# ChatGPT as the Transportation Equity Information Source for Scientific Writing

**Boniphace Kutela, Ph.D., P.E** (Corresponding author)
Texas A&M Transportation Institute
Houston, TX 77024
Email: b-kutela@tti.tamu.edu

**Shoujia Li**
Texas A&M Transportation Institute
Bryan, TX-77807
Email: shoujiali@tamu.edu

**Subasish Das, Ph.D.**
Texas State University
San Marcos, TX-78666
Email: subasish@txstate.edu

**Jinli Liu**
Texas State University
San Marcos, TX-78666
Email: jinli.liu@txstate.edu

**Abstract**

Transportation equity is an interdisciplinary agenda that requires both transportation and social inputs. Traditionally, transportation equity information are sources from public libraries, conferences, televisions, social media, among other. Artificial intelligence (AI) tools including advanced language models such as ChatGPT are becoming favorite information sources. However, their credibility has not been well explored. This study explored the content and usefulness of ChatGPT-generated information related to transportation equity. It utilized 152 papers retrieved through the Web of Science (WoS) repository. The prompt was crafted for ChatGPT to provide an abstract given the title of the paper. The ChatGPT's abstracts were then compared to human-written abstracts using statistical tools and unsupervised text mining. The results indicate that a weak similarity between ChatGPT and human-written abstracts. On average, the human-written abstracts and ChatGPT generated abstracts were about 58% similar, with a maximum and minimum of 97% and 1.4%, respectively. The keywords from the abstracts of papers with over the mean similarity score were more likely to be similar whereas those from below the average score were less likely to be similar. Themes with high similarity scores include access, public transit, and policy, among others. Further, clear differences in the key pattern of clusters for high and low similarity score abstracts was observed. Contrarily, the findings from collocated keywords were inconclusive. The study findings suggest that ChatGPT has the potential to be a source of transportation equity information. However, currently, a great amount of attention is needed before a user can utilize materials from ChatGPT.

Keywords: Transportation equity; Artificial intelligence; ChatGPT

# 1. Introduction

The use of artificial intelligence (AI) is radically altering how individuals live their lives and carry out the many activities they do daily. People's ways of living and working have been revolutionized because of the widespread adoption of cutting-edge technologies such as smartphones and smart watches. In addition, the use of voice command devices like SIRI and Alexa has transformed the day-to-day activities that people participate in.

In November 2022, OpenAI released ChatGPT, an advanced language model, that interacts with users by providing a set of written directives and produces the written text according to the instruction given (ChatGPT & Perlman, 2022; Noever & Ciolino, 2022; Wenzlaff & Spaeth, 2022). This tool has gained the attention of experts in different fields, such as academicians, economists, social scientists, engineers, and computer scientists (ChatGPT & Perlman, 2022; Gao et al., 2022). Most of the concerns surrounding this tool have been about whether ChatGPT will replace some human-generated activities, such as writing codes/algorithms, preparing poems, movie transcripts, etc. (ChatGPT & Perlman, 2022; Qadir, 2022). Some scholars even argue that certain jobs will be replaced by ChatGPT (Qadir, 2022), whereas others disagree by indicating that the tool is not capable of taking over most human-generated jobs and thus will have minimal impact (Aydın & Karaarslan, 2022; Chen & Eger, 2022; Frye, 2022; Wenzlaff & Spaeth, 2022). However, ChatGPT is based on the advanced language model, which uses reinforcement learning, so it is expected to get better when new observations are included in retraining the model.

There has been discussion from researchers and the public about the replacement of Google with ChatGPT (Bindra, 2023; Greene, 2023). Google searches have been used by various researchers as the starting point for searching for various information. A key question among researchers has been whether the future of Google searches is on the brink of collapsing. Due to its ability to create informative texts given a certain prompt, researchers and the public have been using ChatGPT as a source of information. Users have prompted topics related to politics, social science, comedy, business, art, healthcare, games, coding, computer science, and transportation, among others (Kim & Lee, 2023; Yalalov, 2022). In the transportation field, public transportation accessibility and affordability have been among a few areas where users have been searching for information using ChatGPT (Mobility-Innovators, 2023). Public transportation accessibility and affordability form a part of transportation equity, which is among the hot topics both in the United States and globally.

Transportation equity is among the emerging topics of interest in the transportation field. The concept of transportation equity pertains to guaranteeing fair access to transportation alternatives for all individuals, regardless of their socioeconomic position or other individual circumstances (di Ciommo & Shiftan, 2017). With the global trend of urbanization leading to a surge in population and associated problems such as traffic gridlock and environmental pollution, transportation equity is gaining prominence as a pressing issue in various regions. Transportation equity is crucial as it has a pivotal role in advancing social and economic inclusivity(di Ciommo & Shiftan, 2017; Martens et al., 2019). Access to transportation is critical for individuals to access essential services like education, healthcare, and employment opportunities. Inadequate transportation infrastructure or unaffordable transportation options can impede people's ability to access such opportunities(Martens et al., 2019). By enhancing transportation choices and ensuring equal access, transportation equity can contribute to reducing poverty, improving social mobility, and boosting economic development.

Considering the importance of transportation equity, various outlets have been used as sources of information for this topic. Such sources include the National Association of City Transportation Officials (NACTO), the National Center for Mobility Management (NCMM), the American Public Transportation

Association (APTA), and the U.S. Department of Transportation (USDOT). Other sources of information for transportation equity issues include the Web of Science, Google Scholar, and Transportation research in Transport Research International Documentation (TRID) (Clarivate, 2023; TRID, 2023). The information included in these sources include policy papers, design guides, toolkits, webinars, policy briefs, and case studies that address transportation equity issues (APTA, 2023; NACTO, 2023; NCMM, 2023; USDOT, 2023). Information from these sources has been used by researchers and practitioners to address transportation equity issues. Further, researchers utilize such sources for manuscript preparation for publication.

Recently, ChatGPT has been acquired by Microsoft and added to the Bing website, whereby users can chat with it and obtain text-based answers for whatever they are looking for. Like other search engines, ChatGPT can be biased, error-prone, or provide misinformation in its content. There have been some complaints about biased information from ChatGPT in various fields, such as political science, earth science, and social science (Alba, 2023; Bass, 2023; Biddle, 2023). The key difference between ChatGPT compared to other sources is that other search engines, such as Google search, provide the user with multiple sources of the queried information, whereas ChatGPT gives the user a single document with all the information in it. Thus, a user has no opportunity to filter misinforming content. In addition to researchers, the general community relies on search engines to obtain various information. With how easy it is to use ChatGPT, several people might be influenced to use it for day-to-day searches. Thus, the information from this source should be error-free and consistent with other trusted sources to maintain a clear understanding of various issues within the community.

Despite the debate of the biases, misinformation, and errors from ChatGPT, efforts to evaluate the extent of similarity and the content of similar and dissimilar information have still not been done to a great extent. A few studies have attempted to understand the difference between published articles and ChatGPT-generated materials. (Gao et al., 2022; Kutela, Msechu, et al., 2023). These studies concluded that ChatGPT is a good tool for academic writing but needs more human inputs to make the content more human-like. However, these studies did not provide comparable statistical analysis for the two sources of information. In this study, the abstracts of manuscripts were used to explore the difference between human-generated texts in peer-reviewed journals and ChatGPT-generated texts. The intention was to evaluate whether the information from the two sources, human and AI, was similar, and if so, to what extent and what specific content was more similar than others. Further, this study added to the body of literature regarding the methodological approach needed to explore the key difference between human-generated scientific content and ChatGPT texts. It is found that ChatGPT is still not well adept in developing correct citations. In many of the lengthy scientific writing generated by CharGPT generates either non-existing or fake citations. This study thus explores only abstracts (abstracts are citation free) to investigate the authenticity of the contents generated by ChatGPT.

The remaining sections are presented as follows. The next section presents the study methodology and discusses the data description and analytical approaches. The results and discussion section follows, then the conclusion and future studies are presented last.

## 2. Methodology

As described earlier, this study intends to explore whether ChatGPT can produce publication-ready materials comparable to human-generated text in published journals. This section presents the methodological approach used to attain the study objectives. The section is divided into two main sections, data description and analytical methods.

*2.1. Data Description*

To assess the similarities of the information retrieved from different sources, two types of data are necessary, human-written text and ChatGPT-generated text. In this study, authors utilized the abstract section of published papers as the human-written text.

Authors extracted transportation equity papers from the Web of Science database in which "transport equity" and "transportation equity" keywords were used to obtain the transportation equity-related manuscripts. A total of 251 manuscripts containing these keywords in their abstracts were extracted. Upon further pre-processing and duplicate checking, a total of 152 manuscripts were retained for further analysis. The abstract sections, titles, author keywords, and year of publication are the few variables of interest available in the downloaded data. Most of these papers were published in the Transportation Research Record, Transportation Research Part D-Transport and Environment, the Journal of Transport Geography, Transport Policy, and Transportation Research Part A-Policy and Practice.

To obtain the corresponding abstracts generated by ChatGPT, a prompt, which contains the directives of the actions to be taken by ChatGPT, is necessary. The following prompt was used for retrieval:

> *"I want you to develop write an unstructured abstract with minimum of 300 words and maximum of 500 words for publication in a scientific journal that focuses on transport equity. I will give you several titles then I want you to give me the unstructured abstract. You need to adopt a persona of high-profile researcher in transport equity, who has exceptional writing skills and has published over 100 manuscripts from various parts of the world. The abstract should have the details for at least introduction, objectives, methodology, key findings, and study implications. The first title is "Title of the paper."*

The authors supplied all 152 titles to ChatGPT. The corresponding abstracts generated by ChatGPT were extracted and matched to the human-written abstracts in an Excel sheet for further analysis.

*2.2.Analytical Methods*

Three analytical methods, document similarity analysis, text network analysis, and text cluster analysis, were applied to the text data to explore the similarities and content of the transportation equity information from two sources. The titles of the manuscripts, authors, keywords, and abstracts were used for analysis. The document similarity analysis shows the similarity scores between the abstracts generated by ChatGPT and those written by a human. However, it does not describe the content of such abstracts. To explore the similarities and differences in the content, text network analysis and text cluster analysis were applied.

*2.2.1. Document Similarity Analysis*

Document similarity analysis is used to determine the similarities between different documents. Normally, researchers and the public consider documents to be similar if they are semantically close and describe similar concepts/themes (Selivanov, 2018; Zach, 2020).

In this study, the document is defined as the abstract. The comparison is between the abstracts written by humans to the corresponding abstracts written by ChatGPT. For each given title of the study, the human-written abstract and ChatGPT-generated abstracts are extracted and stored in the spreadsheet.

Various approaches, such as Jaccard distance, Cosine distance, Euclidean distance, and Relaxed Word Mover's Distance, can be applied to determine document similarity (Selivanov, 2018; Zach, 2020). This study used Cosine similarity with the LSA approach to determine similarities between two corresponding abstracts.

First, the documents were transformed into bag-of-words to compute the similarity between the two documents, so each document will be a sparse vector. Thus, the similarity between two documents (abstracts) can be computed as:

$$Similarity\ (doc_1, doc_2) = Cos\theta = \frac{doc_1\ doc_2}{|doc_1||doc_2|}$$

Whereby $doc_1$ is the human-written abstract of the paper and $doc_2$ is the ChatGPT-generated abstract of the same paper. The package text2vec (Selivanov, 2022) was used to compute the similarity scores. The similarity score varies between 0 and 1, where a score of 1 implies that the two documents are duplicates and a score of zero implies that the two documents are not similar at all (Zach, 2020).

### 2.2.2. Text Network Analysis

Text Network Analysis (TNA) has been utilized in various fields such as literature and linguistics (Hunter, 2014), traffic safety and operations (Kutela, Das, et al., 2021; Kutela & Teng, 2021; Kwayu et al., 2021), and bibliometrics in transportation studies (Jiang et al., 2020). TNA uses nodes and edges to establish relationships between keywords within a corpus (see **Figure 1**), and its strength lies in its ability to visualize keywords and establish connections among them (Jiang et al., 2020; B. Kutela et al., 2021; Boniphace Kutela et al., 2021; Paranyushkin, 2011). The frequency and co-occurrence of keywords in the network are represented by the size of the nodes and the edges, respectively.

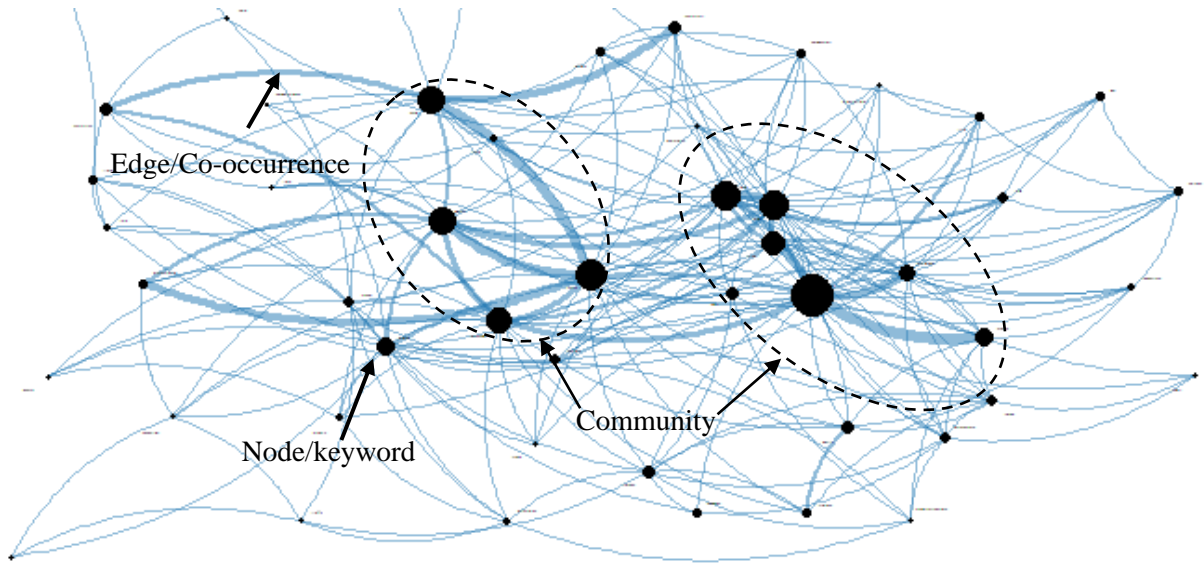

**Figure 1 A Skeleton of the Text Network**

Various processes are performed on the data when conducting TNA analysis. Normalization is the first process, whereby unstructured data is converted to structured data, all symbols are removed, and all texts are converted to lowercase then the output is used to create a matrix of keywords with their frequencies of occurrence (Das & Dutta, 2020; Kutela, Magehema, et al., 2022; Kutela, Novat, et al., 2022). The constructed matrix is then visualized with keywords represented as variously sized nodes based on their recorded frequencies. Various metrics can be utilized for comparative analysis, but document and collocated frequency (Kutela, Combs, et al., 2022; Kutela, Kitali, et al., 2023) are the two metrics used in this study to compare human-generated and ChatGPT-generated introductions. Document frequency is the number of documents that contain the keyword of interest. Keyword frequency, on the other hand, focuses on the number of times the keyword appears in the document. Prior to producing keyword frequency, text stemming was necessary to reduce keywords to their roots/base form. Text stemming allows for better comparison and matching of words with the same root, even if they have different suffixes (Das & Dutta, 2020). For example, "run," "runs," "running," and "runner" can all be stemmed to "run," making it easier to identify all occurrences of the root word. Collocation frequency assesses the number of times the keywords are located next to each other and offers greater insights than individual keywords do since it focuses more on the closeness between two keywords in the corpus. The collocation of keywords in a text network plays a great role in the formation of text clusters, typically referred to as a community of keywords. A community of keywords represents a group collectively clustered in the text network; there can be two or more in a text network (see Figure 1).

### 2.2.3. Text Cluster Analysis

Finally, text clustering using a simple Reinert textual clustering method was used to assess the key clusters for each source of the introduction section (Barnier, 2022). This method uses the same data preparation approach as TNA, but the product is several clusters that represent a certain theme. The results and discussion are presented in the next section.

## 3. Results and Discussion

This section presents the results and discussion. It covers the document similarity results, the text network of the titles and authors keywords, the text network of the abstracts, and the clusters of the abstracts.

### 3.1. Document Similarity Results

**Table 1** presents the similarity scores between human-written and ChatGPT-generated abstracts. It is observed that the abstract is divided into six different metrics used to measure quantitative assessments. The similarity score is measured by how much they are similar on a scale of zero to one. However, some algorithms may produce negative scores, indicating dissimilarity. The similarity score increases from the minimum to the maximum category as more abstracts are developed. For example, there is a notable change between the minimum and the first quartile. On average, the content from human-written and ChatGPT-generated abstracts is 58% similar. The table also shows that the minimum similarity score is 1.4% and the maximum similarity score is nearly 97%. In the statistic, the median is 62%, in which the first quartile is 41% and the third quartile is 80%.

**Table 1 Similarity Scores between Human-Written and ChatGPT-generated Abstracts**

|  | Metrics | | | | | |
| --- | --- | --- | --- | --- | --- | --- |
|  | Minimum | 1st Quartile | Median | Mean | 3rd Quartile | Maximum |
| Similarity Score | 0.01402 | 0.41 | 0.61618 | 0.58428 | 0.79738 | 0.96602 |

Although the document similarity scores present the relationship between the two documents being compared, it does not show the content of the text. Thus, text network analysis was applied to explore the pattern of the key themes. The content analysis involves the titles, author keywords, and abstracts. To perform content analysis, the documents were divided into two parts, the high and low similarity scores. The high similarity score document included all documents with similarity scores above the mean, while the low similarity score document included all documents with mean or lower similarity scores. The next section presents the content analysis results of the titles and author keywords.

*3.2. Text Network Results of the Titles and Author Keywords*

The content of the titles and the authors keywords were evaluated. **Figure** *2* presents the text network of the titles and author's keywords section with high and low similarity scores, while **Error! Reference source not found.** presents the performance metrics of the text networks. The text network of titles with high similarity scores (**Figure** *2* (a)) is heavily centered on the words *accessibility, public, transit, disabilities,* and *urban*. Further, the title sections constitute keywords such as *plan, community, vehicle,* and *mobility*, indicating the abstract also covers mobility of transportation in various studies. Some other keywords have a relatively low representation, including *state, travel, bridge, job,* and *county,* among others.

**Figure** *2* (b) presents the text network for titles with low similarity scores. The text network is heavily centered on the words *accessibility, urban, transit, planning,* and *public,* similar to the titles with high similarity scores. However, the text network nodes of titles with low similarity scores appear to be more decentralized than those presented on the network of titles with high similarity scores. Observing the right side of the title text network for low similarity scores (**Figure** *2*(b)), there is a small network with keywords including *distribute, business, cbd, Sydney,* and *central* observed. These keywords can be related to business in central distribution. In addition, there are some linked words, such as *study* and *case*, *use* and *policy, urban* and *model, research* and *based,* which are related to terms of methodology.

**Figure** *2* (c) presents the text network for the author's keyword sections with high similarity scores. The text network is heavily centered on the words *mobility, policy, urban, accessibility* and *public*. There are some words significantly linked in the author's keywords text network with high similarity scores, including *social* and *communities, shared* and *mobility,* and *urban* and *rural*. These linked keywords show the core of this research.

**Figure** *2* (d) presents the text network for authors keyword sections with low similarity scores. The text network is heavily centered on the keywords *spatial, public, social, travel,* and *accessibility*. However, the text network nodes of authors keywords with low similarity scores also appear to be more decentralized than those presented on the network of authors keywords with high similarity scores. This outcome implies that these keywords with low similarity scores appear to be more varied than those with high similarity scores. Although **Figure** *2* shows that the four networks share some similarities and portray some differences, a comparative analysis of the four networks can be performed using the keyword frequencies. According to the results in **Table** *2* in the group with a high similarity score, among the top 20 keywords, ten are common for both sides. Even for the ten common keywords, the ranking varies significantly. For instance, the keyword *analysis* in title metrics with high similarity scores is ranked 14th, appearing in 7 titles, while it is ranked fifth in the title metrics with low similarity scores, appearing 7 times. Keywords such as *commute, vehicle, China, mobility,* etc., appear only in the title metrics with high similarity scores. On the other hand, the keywords *safety, travel, citi, region,* and *area*, among others, appear only in the title metrics with low similarity scores.

In addition to the title metrics, an analysis of the author's keywords results can be performed. In the author's keywords group with different similarity scores, among the top 20 keywords, there is a slight variation of the keywords and associated rankings. This observation implies that the keywords from high and low similarity score papers do not differ significantly.

**Figure 2 Text Networks of the Content of the Titles and Author's Keywords**

**Table 2 Topmost Keywords from Titles and Author's Keywords**

| | Titles | | | | | | Author Keywords | | | | | |
|---|---|---|---|---|---|---|---|---|---|---|---|---|
| | High Score | | | Low Score | | | High Score | | | Low Score | | |
| Rank | Feature | Freq | Docfreq | Feature | Freq | Docfreq | Feature | Freq | Docfreq | Feature | Freq | Docfreq |
| 1 | transit | 19 | 16 | access | 10 | 10 | access | 18 | 18 | public | 11 | 11 |
| 2 | access | 15 | 14 | transit | 8 | 8 | mobil | 19 | 16 | access | 11 | 9 |
| 3 | plan | 11 | 11 | public | 8 | 8 | public | 14 | 13 | transit | 8 | 6 |
| 4 | urban | 10 | 10 | plan | 8 | 8 | plan | 13 | 13 | urban | 7 | 6 |
| 5 | studi | 10 | 10 | analysi | 7 | 7 | transit | 15 | 12 | social | 5 | 5 |
| 6 | public | 9 | 9 | safeti | 6 | 6 | justic | 12 | 11 | analysi | 5 | 5 |
| 7 | new | 9 | 9 | case | 6 | 6 | social | 10 | 9 | travel | 5 | 5 |
| 8 | commut | 9 | 9 | studi | 6 | 6 | analysi | 10 | 9 | system | 5 | 4 |
| 9 | vehicl | 9 | 8 | urban | 6 | 6 | commut | 10 | 8 | polici | 4 | 4 |
| 10 | impact | 8 | 8 | travel | 5 | 5 | urban | 8 | 8 | spatial | 4 | 4 |
| 11 | use | 8 | 8 | approach | 5 | 5 | sustain | 7 | 7 | mobil | 4 | 4 |
| 12 | mobil | 8 | 8 | citi | 5 | 5 | model | 7 | 6 | model | 3 | 3 |
| 13 | china | 8 | 8 | model | 5 | 4 | vehicl | 7 | 6 | plan | 3 | 3 |
| 14 | analysi | 7 | 7 | practic | 4 | 4 | polici | 6 | 6 | dispar | 2 | 2 |
| 15 | review | 7 | 7 | region | 4 | 4 | share | 7 | 5 | disadvantag | 2 | 2 |
| 16 | case | 7 | 7 | assess | 4 | 4 | environment | 4 | 4 | exclus | 2 | 2 |
| 17 | justic | 6 | 6 | new | 4 | 4 | impact | 4 | 4 | design | 2 | 2 |
| 18 | evid | 6 | 6 | polici | 4 | 3 | choic | 4 | 4 | geographi | 2 | 2 |
| 19 | state | 5 | 5 | research | 3 | 3 | disabl | 5 | 3 | geograph | 2 | 2 |
| 20 | model | 5 | 5 | area | 3 | 3 | travel | 5 | 3 | inform | 2 | 2 |

Note: Docfreq = is the document frequency, which is the number of documents

### 3.3. Text Network Results of the Abstracts

The content of the abstracts written by humans and by ChatGPT was evaluated. **Figure 3** presents the text network of the human-written abstract section and ChatGPT-generated abstract section with high and low similarity scores, while **Table 3** and **Table 4** present the performance metrics of the text networks. The text network of human-written abstracts with high similarity scores (**Figure 3** (a)) is heavily centered on the keywords *accessibility, public, transit, travel,* and *planning*. This is because the studies used in this paper are transportation-related. Further, the human-written abstract sections constitute keywords such as *plan, safety, environmental,* and *travel*, indicating the abstract also covers travel concerns in various studies. Some other keywords have a relatively low representation, including *time, regional, commuting, mobility,* and *services,* among others.

**Figure 3** (b) presents the text network for ChatGPT-generated abstracts with high similarity scores. The text network is heavily centered on the keywords *accessibility, urban, transit, plan,* and *public,* similar to the human-written text network. However, the human-written text network nodes appear to be larger than those presented on the ChatGPT network. This outcome implies that these keywords appear more frequently in human-written abstract sections than in the ChatGPT-generated ones. Observing the right side of the ChatGPT-generated text network with high similarity scores (**Figure 3** (b)), the linked keywords include *accessibility* and *key, low* and *income, communities* and *income,* and *public* and *transit*. These linked keywords can be related to transportation accessibility and the impact of income on transportation. On the other hand, the left side of the text network contains linked keywords such as *key* and *finding*, *survey* and *data, data* and *analysis, literature* and *review, data* and *used,* and *survey* and *methods,* which are related to analysis and recommendations/findings.

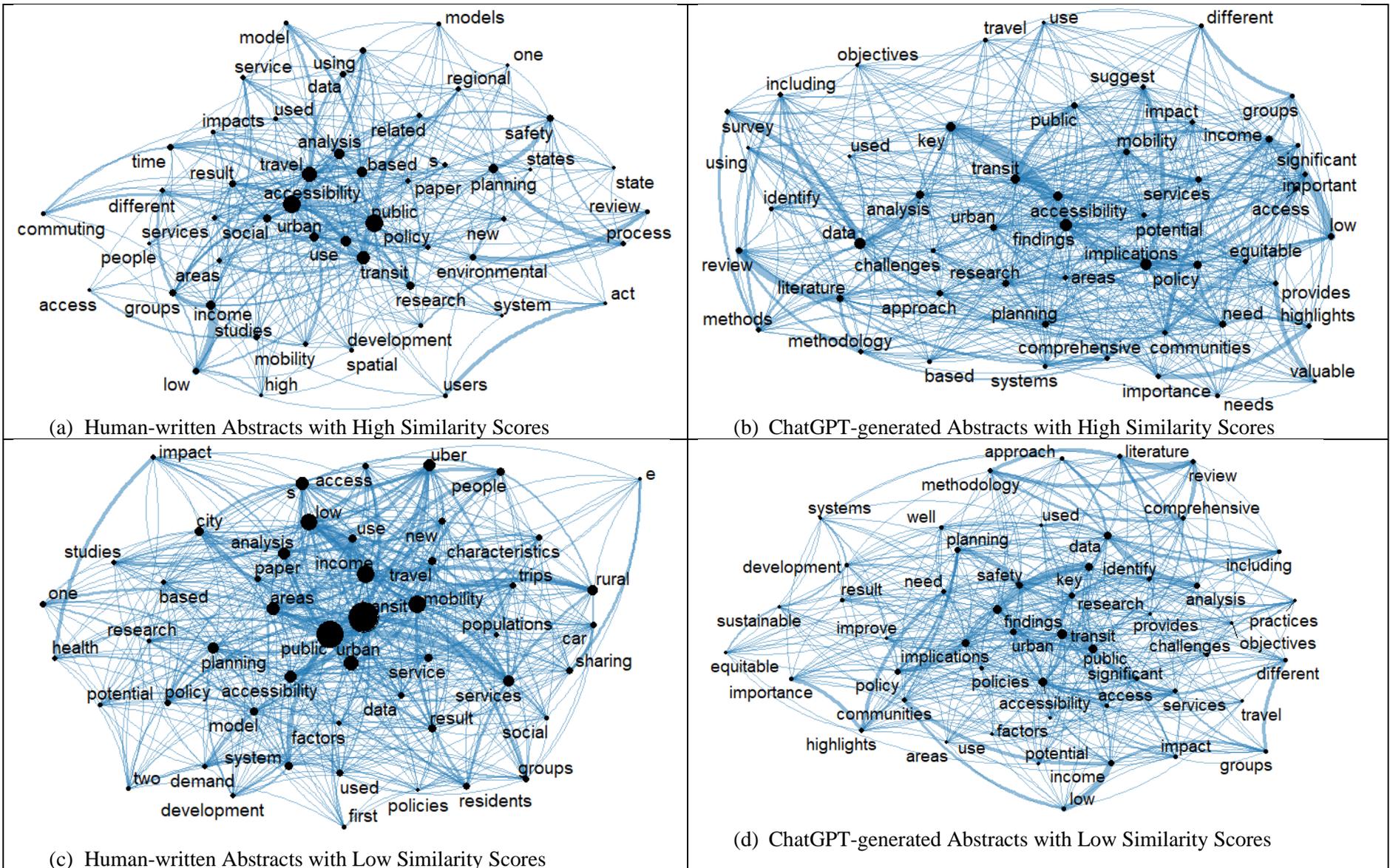

**Figure 3 Text Networks of the Content of the Abstracts**

**Table 3 Topmost Keywords from Abstracts with High Similarity Scores**

| Keywords | | | | | | Collocations | | | |
|---|---|---|---|---|---|---|---|---|---|
| GPT | | | Human | | | GPT | | Human | |
| Feature | Freq | Docfreq | Feature | Freq | Docfreq | Collocation | Count | Collocation | Count |
| access | 431 | 113 | access | 225 | 67 | key findings | 139 | public transit | 39 |
| use | 288 | 114 | use | 212 | 105 | highlights importance | 55 | 21st century | 25 |
| find | 274 | 151 | public | 211 | 69 | public transit | 66 | act 21st | 23 |
| need | 258 | 117 | transit | 187 | 51 | provides valuable | 59 | planning process | 15 |
| includ | 249 | 129 | plan | 149 | 64 | literature review | 49 | flexible efficient | 15 |
| provid | 245 | 124 | travel | 145 | 51 | valuable insights | 49 | new york | 15 |
| polici | 233 | 106 | servic | 118 | 46 | highlights need | 39 | accountable flexible | 15 |
| data | 224 | 112 | polici | 110 | 63 | case studies | 28 | safe accountable | 15 |
| transit | 216 | 38 | urban | 107 | 52 | importance considering | 51 | efficient act | 14 |
| implic | 215 | 148 | model | 106 | 44 | qualitative quantitative | 27 | legacy users | 14 |
| plan | 203 | 67 | develop | 106 | 66 | current state | 26 | act legacy | 14 |
| public | 202 | 63 | analysi | 106 | 60 | conclusion provides | 32 | land use | 14 |
| key | 196 | 141 | research | 100 | 58 | findings indicate | 66 | job accessibility | 13 |
| urban | 192 | 64 | system | 100 | 54 | important implications | 35 | disadvantaged groups | 12 |
| import | 189 | 117 | area | 94 | 51 | implications suggest | 34 | travel demand | 12 |
| communiti | 188 | 61 | result | 88 | 70 | methodology includes | 26 | public services | 12 |
| improv | 187 | 81 | group | 85 | 48 | inform development | 24 | environmental review | 11 |
| research | 186 | 89 | studi | 85 | 37 | population groups | 24 | york city | 10 |
| servic | 186 | 54 | social | 84 | 44 | policies programs | 25 | autonomous vehicles | 10 |
| analysi | 178 | 96 | mobil | 83 | 38 | review existing | 24 | united states | 10 |

**Table 4 Topmost Keywords from Abstracts with Low Similarity Scores**

| Keywords | | | | | | Collocations | | | |
|---|---|---|---|---|---|---|---|---|---|
| GPT | | | Human | | | GPT | | Human | |
| Feature | Freq | Docfreq | Feature | Freq | Docfreq | Collocation | Count | Collocation | Count |
| find | 98 | 55 | use | 76 | 37 | key findings | 50 | low income | 24 |
| implic | 77 | 54 | polici | 45 | 30 | public transit | 33 | public transit | 22 |
| object | 67 | 53 | public | 86 | 28 | low income | 33 | e scooters | 9 |
| key | 74 | 52 | result | 35 | 28 | findings indicate | 27 | land use | 9 |
| includ | 88 | 48 | transit | 110 | 27 | united states | 25 | subway accessibility | 8 |
| need | 89 | 47 | develop | 42 | 27 | provides valuable | 23 | public services | 8 |
| provid | 92 | 46 | citi | 41 | 27 | literature review | 21 | 95 ci | 8 |
| data | 82 | 43 | analysi | 45 | 26 | best practices | 20 | irr 95 | 8 |
| import | 66 | 43 | access | 63 | 24 | highlights importance | 19 | rural areas | 7 |
| use | 100 | 42 | servic | 56 | 22 | highlights need | 19 | lower income | 7 |
| highlight | 57 | 42 | incom | 39 | 21 | valuable insights | 19 | planning processes | 7 |
| access | 163 | 40 | improv | 32 | 21 | importance considering | 18 | bike sharing | 7 |
| polici | 106 | 40 | provid | 24 | 21 | ride hail | 18 | new york | 7 |
| signific | 56 | 38 | plan | 62 | 20 | road safety | 17 | transit pass | 7 |
| analysi | 72 | 37 | mobil | 47 | 20 | marginalized communities | 16 | last mile | 7 |
| methodolog | 43 | 37 | data | 26 | 20 | current state | 15 | act 21st | 7 |
| develop | 59 | 36 | system | 39 | 19 | conclusion provides | 15 | 21st century | 7 |
| literatur | 48 | 36 | paper | 23 | 19 | mass transit | 15 | shared mobility | 6 |
| research | 73 | 34 | urban | 41 | 18 | mixed methods | 14 | york city | 6 |
| studi | 40 | 34 | area | 39 | 18 | public services | 14 | first last | 6 |

Note: Docfreq = is the document frequency, which is the number of documents

**Figure 3** (c) presents the text network for human-written abstracts with low similarity scores. The text network is heavily centered on the keywords *transit, public, income, mobility,* and *access,* similar to the text network with high similarity scores. However, the text network nodes appear to be larger than those presented on the human-written network with high similarity scores. This outcome implies that these keywords appear more frequently in human-written sections with higher similarity scores than in the sections with low similarity scores. The methodology keywords, *research, analysis,* and *model,* are also presented in this section.

**Figure 3** (d) presents the text network for ChatGPT-generated abstracts with low similarity scores. The text network is heavily centered on the keywords *safety, transit, urban, planning* and *public,* similar to the human-written text network. Observing the left side of the ChatGPT-generated text network with high similarity scores (**Figure 3** (d)), the linked keywords include *low* and *income, communities* and *income,* and *public* and *transit.* These linked keywords are similar to the text network for ChatGPT-generated abstracts with high similarity scores. On the other hand, the right side of the text network also contains linked keywords such as *key* and *finding*, *identify* and *data, data* and *analysis, literature* and *review, approach* and *used,* and *data* and *used,* which are related to terms of analysis and recommendations/findings.

Although **Figure 3** shows that the four networks share some similarities and portray some differences, a comparative analysis of the four networks can be performed using the keyword and collocation frequencies. According to the results in **Table 3** in the group with high similarity scores, among the top 20 keywords, ten are common for both sides. Even for the ten common keywords, the ranking varies significantly. For instance, the keyword *transit* in ChatGPT-generated metrics is ranked ninth, appearing in 216 abstracts, while it is ranked fourth in the human-written metrics, appearing 187 times. Keywords such as *travel, model, system, group,* etc., appear only in the human-written metrics. On the other hand, the keywords *find, need, data, key,* and *community*, among others, appear only in the ChatGPT-generated metrics.

**Table 4** shows similar results to the metrics generated by ChatGPT and humans with low similarity scores. There are only seven keywords common for both sides with significantly different ranks, like *polici* which is ranked third in human-written abstracts, appearing 86 times, while only ranked 13[th] in ChatGPT-generated abstracts, appearing 106 times. Keywords such as *find, object, key* and *include*, among others, appear only in ChatGPT-generated metrics. Keywords *result, income, transit,* and *improve*, among others, appear only in human-written metrics. This observation indicates that ChatGPT is capable of replicating some keywords.

In addition to the individual keywords, the collocated keywords results can be used to distinguish ChatGPT-generated abstracts from human-written abstracts. The results in **Table 3** and **Table 4** show that the two approaches differ significantly. In **Table 3**, which shows the metrics with high similarity scores, only one pair of collocated keywords, *public transit*, is common for both approaches. According to the results in **Table 4**, there are three pairs of collocated keywords that are common for both approaches with low similarity scores, *public transit, low income* and *public services.* The keywords associated with the use of the study can be observed in the collocated keywords. Such keywords include *key findings, highlight importance, conclusion provides, findings indicate,* etc. These findings indicate that the ChatGPT algorithm tends to summarize the conclusions and the possible use of the research, while that is not the case for most human-written scientific writeups. The text network and associated metrics provide the details of the content of the abstracts, which facilitates the comparison of the two.

*3.4. Text Clusters of the Abstracts*

**Figure 4** presents the six clusters of ChatGPT-generated and human-written abstracts with high or low similarity scores. According to the clusters for human-written abstracts with high similarity scores (**Figure 4** (a)), the first cluster contains keywords such as *accessibility, social, mobility, system, urban,* and *transit*. Such keywords can be associated with transportation accessibility and mobility systems in urban areas. The second cluster presents keywords related to the analysis of accessibility in terms of travel. Such keywords include *research, individual, travel, time,* and *analysis*. Most of the keywords in the third cluster, such as *accessibility, income, low, groups,* and *high,* can be associated with the relationship between income and accessibility in transport studies. Keywords referring to research on public transit commuting and the factors that influence it can be observed in the fourth cluster, with keywords such as *public, commuting, housing, time,* and *residents*. The fifth cluster produced from the human-written abstracts contains keywords such as *evidence, justice, methods, approach,* and *outcomes*. These keywords relate to evidence-based methods to analyze impact. The last cluster, cluster number six, contains keywords about the Transportation Equity Act, such as *act, state, management, legacy, public, program,* and *century.*

**Figure 4** (b) presents clusters derived from ChatGPT-generated abstracts with high similarity scores. From the figure, these keywords may be related to income and accessibility of public transport services in urban areas; such keywords are *income, findings, access, low, urban,* and *accessibility.* On the other hand, the second cluster contains keywords such as *making, importance, highlights, decision, challenges,* and *research.* These keywords are linked to decision-making processes.

Keywords relating to implications of policy and planning decisions can be observed in the third cluster; the keywords include *policy, implications, conclusion, policymaker,* and *importance*. The fourth cluster includes keywords such as *bridges, development, objectives, growing, traffic,* and *findings;* such keywords explain the development of transportation infrastructure. The fifty cluster keywords are related to the methods used to gather and analyze data, and keywords such as *data, analysis, methods, methodology, approach,* and *used* can be observed. The final cluster, cluster six, contains keywords such as *review, literature, methodology, existing, studies, data,* and *case*; these keywords are associated with literature reviews.

**Figure 4** (c) presents clusters derived from human-written abstracts with low similarity scores. The figure shows that there are some keywords in the first cluster like *Uber*, *mobility*, *city*, *areas*, *urban,* and *trips*, which are related to public services and the impact of new mobility services such as Uber. In the second cluster, the keywords are associated with the analysis of existing work about accessibility, with keywords such as *accessibility*, *research*, *analysis*, *work*, *studies*, *existing,* and *paper* included in this cluster. Keywords in the third cluster like *service*, *demand*, *car*, *users*, *trips,* and *socio* show that this cluster is mainly about the analysis of car sharing services. The fourth cluster contains keywords like *mobility*, *time*, *low*, *income,* and *services*, which are related to the transportation and mobility of low-income population. The keywords in the fifth cluster include *public*, *use*, *used*, *safety*, *model*, *system*, *environment,* and *impacts*, benefits which are likely associated with the use of public transit systems and their impacts on the environment. The last cluster includes *urban*, *rural*, *groups*, *residents*, and *services*, which shows the cluster is associated with public transportation services in urban and rural areas.

**Figure 4** (d) presents clusters derived from ChatGPT-generated abstracts with low similarity scores. The figure shows that there are some keywords in the first cluster that are mainly associated with research methods and data analysis in the field of urban transit and accessibility, like *quantitative*, *data*, *methods*, *qualitative*, *analysis,* and *findings*. In the second cluster, the keywords are associated with the accessibility for different income groups in urban areas, with *low*, *income*, *different*, *spatial*, *area*, *population,* and *accessibility* included in this cluster. Keywords in the third cluster, such as *review*, *literature*, *interviews*, *case*, *analysis,* and *studies,* show that this cluster is mainly about data analysis and literature review. The fourth cluster contains keywords like *low*, *income*, *implications*, *population,* and *impact*, which are related to the impact of transportation systems on communities, particularly low-income populations. The keywords in the fifth cluster, *safety*, *road*, *improve*, *policies*, *findings*, *services,* and *potential,* are likely associated with road safety and the improvements needed in the field. The last cluster includes *policymakers*, *provide*, *valuable*, *insights*, *planner,* and *need*, which shows that the cluster is associated with the potential methods to help policymakers to improve transportation.

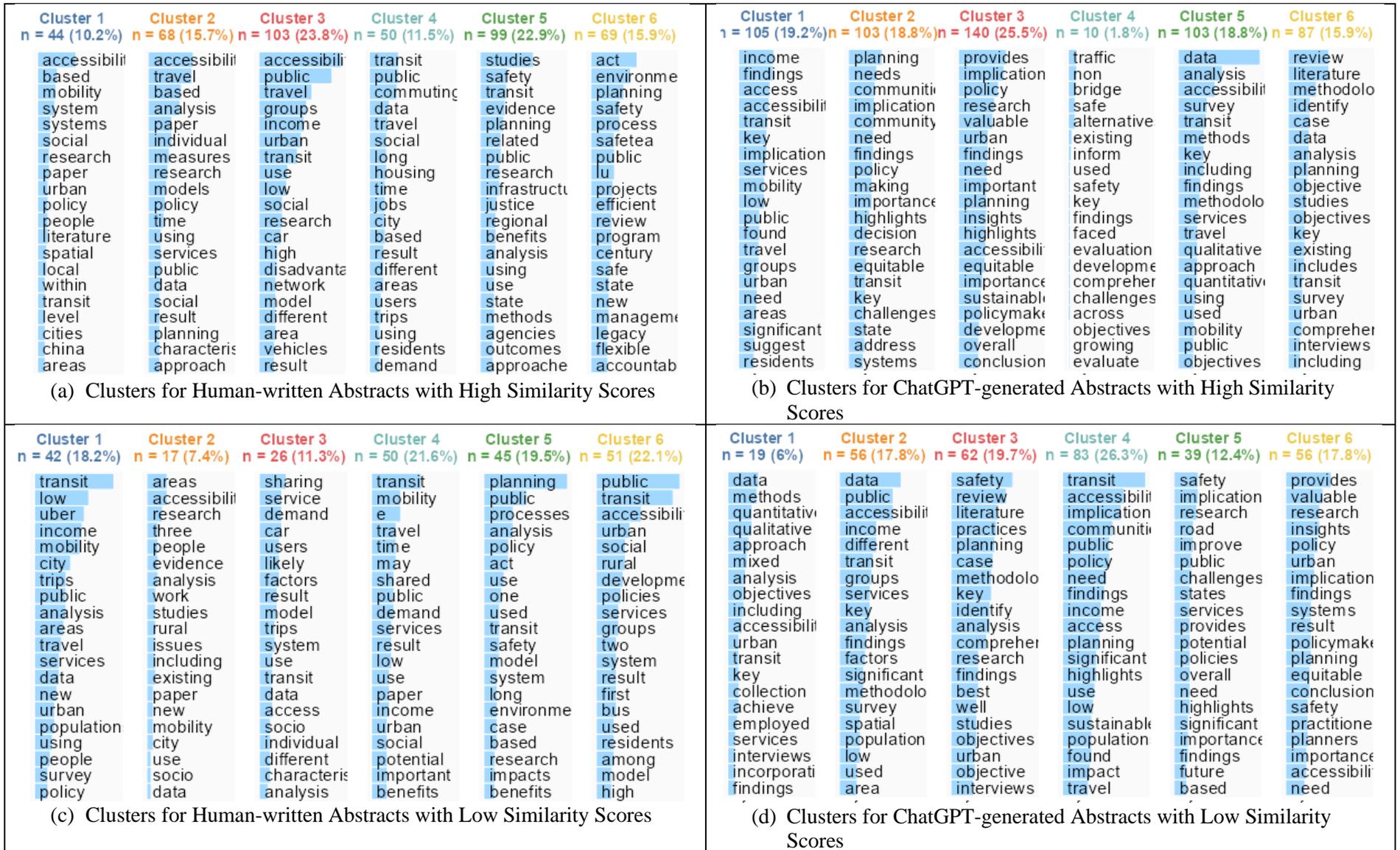

Figure 4 Clusters of the Abstracts

Some similarities and differences can be observed from the clusters generated by ChatGPT and human-written abstracts. For example, similarities can be observed between keywords related to the relationship between income and accessibility in transport studies (ChatGPT's cluster with high similarity scores in Cluster 1, human-written abstracts with high similarity scores in Cluster 3, ChatGPT's cluster with low similarity scores in Clusters 2 & 4, and human-written abstracts with high similarity scores in Cluster 4), using common keywords such as *low*, *income*, *public*, *accessibility*, *high*, *services*, and *social*. Further, there are similarities in data analysis and methodology (ChatGPT's cluster with high similarity scores in Cluster 5 & 6, human-written abstracts with high similarity scores in Cluster 5, ChatGPT's clusters with low similarity scores in Cluster 1, and human-written abstracts with high similarity scores in Cluster 3), common keywords observed in the clusters are *data*, *methods*, *analysis*, and *methodology*.

In addition, the keywords about decision-making and improvement are similar in ChatGPT-generated abstracts with high similarity scores (Clusters 2 & 3) and ChatGPT-generated with low similarity scores (Cluster 6). The keywords which are related to literature reviews are common in ChatGPT-generated abstracts with high similarity scores (Cluster 6), in ChatGPT-generated abstracts with low similarity scores (Cluster 3), and in human-written abstracts with low similarity scores (Cluster 2).

Conversely, differences in the generated abstracts are noted. For example, human-written keywords with high similarity scores focused more on public transit commuting (Cluster 4), accessibility in terms of travel (Cluster 2), and the Transportation Equity Act (Cluster 6). On the other hand, keywords from ChatGPT-generated clusters with high similarity scores focused on the development of transportation infrastructure (Cluster 4). In addition, keywords from human-written abstracts with low similarity scores focused on public services and the impact of new mobility services (Cluster 1), the use of public transit systems and their impacts on the environment (Cluster 5), and public transportation services in urban and rural areas (Cluster 6). Road safety and the improvements needed are mainly focused on by the ChatGPT-generated abstracts with low similarity scores (Cluster 5).

## 4. Conclusions and Future Studies

This study presents a comparative analysis of transportation equity-related themes by considering the source of information. The two sources compared in this study are the traditional source, which is human-written materials, and AI-generated materials. A total of 152 highly cited papers in transportation equity were extracted from the Web of Science. A prompt was then prepared and supplied to ChatGPT to produce abstracts given the title of the manuscripts. Document similarity analysis, text network analysis, and text cluster analysis were applied to determine the relationship between human-written and ChatGPT-generated content.

It was found that on average, the human-written and ChatGPT-generated content are about 54% similar, with a minimum score of 1.4% and a maximum score of 96.7%. The content analysis for high similarity score content (above average) and low similarity score content (below average) was for the titles, authors, keywords, and abstracts. The key themes from the title and author keywords did not differ significantly across the lower and higher similarity scores. On the other hand, there was observed a significant difference in the content of the abstract. In this case, the ChatGPT-generated materials appear to be more generic, especially when the collocation metric is considered. Furthermore, the cluster analysis results revealed significantly different patterns between human-written and ChatGPT-generated abstracts.

Based on these findings, this study concludes that at this moment the transportation equity information retrieved through ChatGPT greatly differs from the human-written content. Therefore, researchers and the general community should exercise care when retrieving transportation equity information from ChatGPT. It is advised that researchers perform a comprehensive review of the information retrieved through ChatGPT with other human-written information. Although the two sources differ significantly, there were observed a few cases where the resulting content had high similarity scores. This is an indication that there are a few topics within transportation equity that ChatGPT performed relatively better. Thus, for such topics, the information retrieved can be used even without conducting a comprehensive review.

Although this study successfully showed the similarities and differences between the information retrieved through ChatGPT against the one written by humans, several limitations exist. First, this study utilized transportation equity-related studies. These studies are not very common for the general community, thus, the chance that their key themes were used to train ChatGPT is relatively lower. That being the case, retrieving such information using ChatGPT is relatively lower. Future studies may consider studies in areas that are common to the general community, such as social science studies. Further, this study considered the abstract section of the published papers. Abstracts tend to contain various information including the objective, methodology, and key findings. It is relatively difficult for even a human to provide an abstract when given a title. Future studies may focus on other easily generatable parts of the manuscripts given a title, such as the study objectives.

## 5. Funding

This study received no funding of any type.

## 6. References

Alba, D. (2023). *ChatGPT, Open AI's Chatbot, Is Spitting Out Biased, Sexist Results - Bloomberg*. Bloomberg. https://www.bloomberg.com/news/newsletters/2022-12-08/chatgpt-open-ai-s-chatbot-is-spitting-out-biased-sexist-results#xj4y7vzkg

APTA. (2023). *Home - American Public Transportation Association*. https://www.apta.com/

Aydın, Ö., & Karaarslan, E. (2022). OpenAI ChatGPT Generated Literature Review: Digital Twin in Healthcare. *SSRN Electronic Journal*. https://doi.org/10.2139/SSRN.4308687

Barnier, J. (2022). *Introduction to rainette*. https://juba.github.io/rainette/articles/introduction_en.html#double-clustering

Bass, D. (2023). *ChatGPT maker OpenAI promises to fix the chatbot's biases | Fortune*. https://fortune.com/2023/02/16/chatgpt-openai-bias-inaccuracies-bad-behavior-microsoft/

Biddle, S. (2023). *The Internet's New Favorite AI Proposes Torturing Iranians and Surveilling Mosques*. The Concept. https://theintercept.com/2022/12/08/openai-chatgpt-ai-bias-ethics/

Bindra, J. (2023). *Will ChatGPT replace Google as our go-to web search platform? | Mint*. https://www.livemint.com/opinion/columns/will-chatgpt-replace-google-asour-go-to-web-search-platform-11671733523981.html

ChatGPT, O. A. A., & Perlman, A. (2022). The Implications of OpenAI's Assistant for Legal Services and Society. *SSRN Electronic Journal*. https://doi.org/10.2139/SSRN.4294197

Chen, Y., & Eger, S. (2022). Transformers Go for the LOLs: Generating (Humourous) Titles from Scientific Abstracts End-to-End. *ArXiv*. https://doi.org/10.48550/arxiv.2212.10522

Clarivate. (2023). *Web of Science Platform - Web of Science Group*. https://clarivate.com/webofsciencegroup/solutions/webofscience-platform/

Das, S., & Dutta, A. (2020). Characterizing public emotions and sentiments in COVID-19 environment: A case study of India. *Https://Doi.Org/10.1080/10911359.2020.1781015*, 1–14. https://doi.org/10.1080/10911359.2020.1781015

di Ciommo, F., & Shiftan, Y. (2017). Transport equity analysis. *Transport Reviews*, *37*(2), 139–151. https://doi.org/10.1080/01441647.2017.1278647

Frye, B. L. (2022). *Should Using an AI Text Generator to Produce Academic Writing Be Plagiarism?* https://papers.ssrn.com/abstract=4292283

Gao, C. A., Howard, F. M., Markov, N. S., Dyer, E. C., Ramesh, S., Luo, Y., & Pearson, A. T. (2022). Comparing scientific abstracts generated by ChatGPT to original abstracts using an artificial intelligence output detector, plagiarism detector, and blinded human reviewers. *BioRxiv*, 2022.12.23.521610. https://doi.org/10.1101/2022.12.23.521610

Greene, T. (2023). *ChatGPT will not replace Google Search | Impact of Social Sciences*. https://blogs.lse.ac.uk/impactofsocialsciences/2023/01/27/chatgpt-will-not-replace-google-search/

Hunter, S. (2014). A Novel Method of Network Text Analysis. *Open Journal of Modern Linguistics*, *4*, 350–366. https://doi.org/10.4236/ojml.2014.42028

Jiang, C., Bhat, C. R., & Lam, W. H. K. (2020). A bibliometric overview of Transportation Research Part B: Methodological in the past forty years (1979–2019). *Transportation Research Part B: Methodological*, *138*, 268–291. https://doi.org/10.1016/j.trb.2020.05.016

Kim, J., & Lee, J. (2023). How does ChatGPT introduce transportation problems and solutions in North America? *SSRN Electronic Journal*. https://doi.org/10.2139/SSRN.4349774

Kutela, B., Combs, T., John Mwekh'iga, R., & Langa, N. (2022). Insights into the long-term effects of COVID-19 responses on transportation facilities. *Transportation Research Part D: Transport and Environment*, *111*, 103463. https://doi.org/10.1016/J.TRD.2022.103463

Kutela, B., Das, S., & Dadashova, B. (2021). Mining patterns of autonomous vehicle crashes involving vulnerable road users to understand the associated factors. *Accident Analysis & Prevention*, 106473. https://doi.org/10.1016/J.AAP.2021.106473

Kutela, B., Kitali, A. E., Kidando, E., Langa, N., Novat, N., & Mwende, S. (2023). Exploring commonalities and disparities of seattle residents' perceptions on dockless bike-sharing across gender. *City, Culture and Society*, *32*, 100503. https://doi.org/10.1016/J.CCS.2023.100503

Kutela, B., Langa, N., Mwende, S., Kidando, E., Kitali, A. E., & Bansal, P. (2021). A text mining approach to elicit public perception of bike-sharing systems. *Travel Behaviour and Society*, *24*, 113–123. https://doi.org/10.1016/j.tbs.2021.03.002

Kutela, B., Magehema, R. T., Langa, N., Steven, F., & Mwekh'iga, R. J. (2022). A comparative analysis of followers' engagements on bilingual tweets using regression-text mining approach. A case of Tanzanian-based airlines. *International Journal of Information Management Data Insights*, *2*(2), 100123. https://doi.org/10.1016/J.JJIMEI.2022.100123

Kutela, B., Msechu, K., Das, S., & Kidando, E. (2023). Chatgpt's Scientific Writings: A Case Study on Traffic Safety. *SSRN Electronic Journal*. https://doi.org/10.2139/SSRN.4329120

Kutela, B., Novat, N., Adanu, E. K., Kidando, E., & Langa, N. (2022). Analysis of residents' stated preferences of shared micro-mobility devices using regression-text mining approach. *Transportation Planning and Technology*, *45*(2), 159–178. https://doi.org/10.1080/03081060.2022.2089145

Kutela, B., Novat, N., & Langa, N. (2021). Exploring geographical distribution of transportation research themes related to COVID-19 using text network approach. *Sustainable Cities and Society*, *67*. https://doi.org/10.1016/j.scs.2021.102729

Kutela, B., & Teng, H. (2021). The Use of Dynamic Message Signs (DMSs) on the Freeways: An Empirical Analysis of DMSs Logs and Survey Data. *Journal of Transportation Technologies*, *11*(01), 90–107. https://doi.org/10.4236/jtts.2021.111006

Kwayu, K. M., Kwigizile, V., Lee, K., Oh, J.-S., & Oh, J.-S. (2021). Discovering latent themes in traffic fatal crash narratives using text mining analytics and network topology. *Accident Analysis and Prevention*, *150*, 105899. https://doi.org/10.1016/j.aap.2020.105899

Martens, K., Bastiaanssen, J., & Lucas, K. (2019). Measuring transport equity: Key components, framings and metrics. *Measuring Transport Equity*, 13–36. https://doi.org/10.1016/B978-0-12-814818-1.00002-0

Mobility-Innovators. (2023). *How ChatGPT will be game changer for Public Transit agencies? - Mobility Innovators*. https://mobility-innovators.com/how-chatgpt-will-be-game-changer-for-public-transit-agencies/

NACTO. (2023). *National Association of City Transportation Officials | National Association of City Transportation Officials*. https://nacto.org/

NCMM. (2023). *Home - National Center for Mobility Management*. https://nationalcenterformobilitymanagement.org/

Noever, D., & Ciolino, M. (2022). The Turing Deception. *ArXiv*. https://doi.org/10.48550/arxiv.2212.06721

Paranyushkin, D. (2011). Identifying the Pathways for Meaning Circulation using Text Network Analysis. *Venture Fiction Practices*, *2*(4). www.noduslabs.com

Qadir, J. (2022). *Engineering Education in the Era of ChatGPT: Promise and Pitfalls of Generative AI for Education*. https://doi.org/10.36227/TECHRXIV.21789434.V1

Selivanov, D. (2018). *Documents similarity*. https://text2vec.org/similarity.html

Selivanov, D. (2022). *text2vec: Modern Text Mining Framework for R*. Comprehensive R Archive Network (CRAN). https://CRAN.R-project.org/package=text2vec

TRID. (2023). *About TRID | Information Services*. https://www.trb.org/InformationServices/AboutTRID.aspx

USDOT. (2023). *Equity | US Department of Transportation*. https://www.transportation.gov/priorities/equity

Wenzlaff, K., & Spaeth, S. (2022). Smarter than Humans? Validating how OpenAI's ChatGPT Model Explains Crowdfunding, Alternative Finance and Community Finance. *SSRN Electronic Journal*. https://doi.org/10.2139/SSRN.4302443

Yalalov, D. (2022). *100 Best ChatGPT Prompts to Unleash AI's Potential | Metaverse Post*. https://mpost.io/100-best-chatgpt-prompts-to-unleash-ais-potential/

Zach. (2020). *How to Calculate Jaccard Similarity in R*. https://www.statology.org/jaccard-similarity-in-r/